# Silicon tracking DAQ

Aurore Savoy-Navarro[1], Alex Charpy[1], Catalin Ciobanu[1], Jacques David[1], Marc Dhellot[1], Jean F. Genat[1], Th. Hung Pham[1], Rachid Sefri[1], Raimon Casanova[2], Albert Comerma[2], Angel Dieguez[2], David Gascon[2] *

1 – LPNHE – Université Pierre et Marie Curie/IN2P3-CNRS
4, Place Jussieu, 75252 Paris-Cedex05 – France

2 – Universitat de Barcelona – Department d'Estructura i Constituents de la Materia,
Avenida Diagonal, Barcelona – Spain

Some preliminary thoughts on how to design and develop the DAQ architecture for the Silicon tracking system at the future Linear electron positron collider, are briefly presented here. The proposed structure includes three DAQ levels. The first level is based on a high level processing mix-mode ASIC sitting on the detector. The second level still on the detector is a DSP like interface that will send the processed data to the general DAQ system. Several novel technological aspects are part of this development. The role of the ongoing test beam activities with detector prototypes as training camp is emphasized.

## 1 Introduction

The SiLC (Silicon tracking for the Linear Collider) R&D collaboration [1] has recently started to study the next Electronics R&D steps, beyond the Front End and Readout Electronics, and how to integrate them within the overall DAQ architecture of the future Linear collider experiments. The proposed DAQ flow chart is presented here in a still very preliminary status, and its main components (levels) are briefly discussed. The role of the ongoing test beam activities as training camp for developing these future DAQ systems is emphasized together with the need of high tech developments in collaboration with Industry.

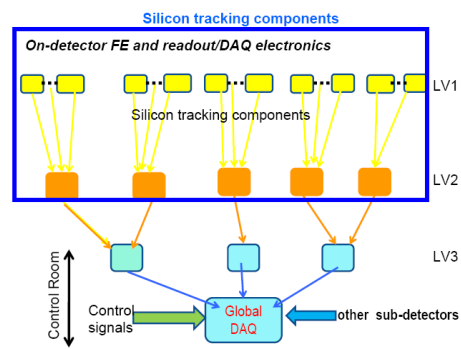

**Figure 1:** Si tracking DAQ flow diagram

## 2 Silicon tracking DAQ architecture and main parameters

The proposed DAQ for the Silicon tracking [2] is foreseen into three main levels, two of them being on the detector. The overall DAQ flow diagram is sketched on Figure 1. Before describing it further let's remind some basic parameters that are driving the overall Silicon DAQ design. First, the proposed DAQ framework applies to all the ILC detector concepts with Silicon tracking components (i.e. SiD, ILD and now also 4[th] concept). The number of Silicon tracking channels to be read out is of the order of several millions (ex: $10^6$ channels in

* This work is performed in collaboration with several SiLC R&D Institutes [1] and their contribution is acknowledged here as well as the partial financial and infrastructure support of the EUDET E.U. project.

the ILD case and of the same order for SiD). The zero suppression is mandatory especially for the components which, due to their location, have a very low occupancy.

Another basic important parameter is of course the machine cycle. It drives the functioning of the overall electronics readout and DAQ (Figure 2). In the ILC case, there is beam during 1ms followed by 200 ms without beam. The DAQ operation will take advantage of this machine cycle to perform on detector some more advanced data processing after digitization of the signals. The time during beam is on will be used for a rather sophisticated pulse height reconstruction, data storage and zero suppression. Even if the time stamping is not a prerequisite as in the CLIC case, the designed FE electronics for Silicon tracking include an electronic time stamping; this is essential to properly process the Silicon data and also useful for other sub-detectors. The time in between bunch train will be used for A/D conversion, power cycling and eventually some calibration.

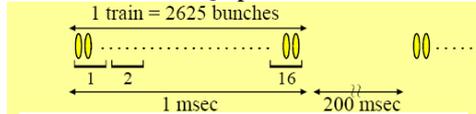

**Figure 2:** ILC beam crossing structure

## 3   The Silicon tracking DAQ levels

The presently foreseen architecture includes the following three levels.

### 3.1   The Level 1-DAQ: the FEE chip

The FEE chip is the crucial piece in the overall DAQ architecture. It does it almost all by itself, thanks to the high degree of signal processing included in this full custom ASIC. Figure 3 shows the layout and photograph of the present prototype described at this workshop [3].

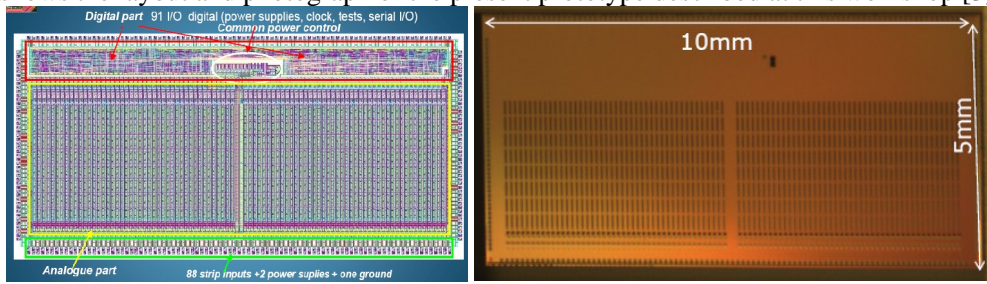

**Figure 3:** layout and photograph of the new SiTR_130-88 chip

This prototype made in 130nm CMOS technology includes almost all the functions of the final F.E ASIC, i.e. a low noise preamplifier with a rather slow shaper (shaping time between 0.5 to 1μs typically), zero suppression and a sophisticated 2D memory structure with 8x8 channels both for pulse height reconstruction and storage and a Wilkinson A/D converter with a high multiplexing factor [4]. Special attention is given to achieve low noise and low power dissipation with this analog chain. The chip functioning is fully digital controlled; All the parameters such as bias voltage (10 bits) and current (8 bits), sparsifier thresholds, event tag and time tag generation, internal calibration system, shaping time frequency and sampling frequencies are all programmable. This ensures a high degree of fault tolerance and high flexibility and robustness. The chip can be triggered internally (sparsifier) or by LVTTL trigger levels [4]. The signal processing is achieved during acquisition (beam on).

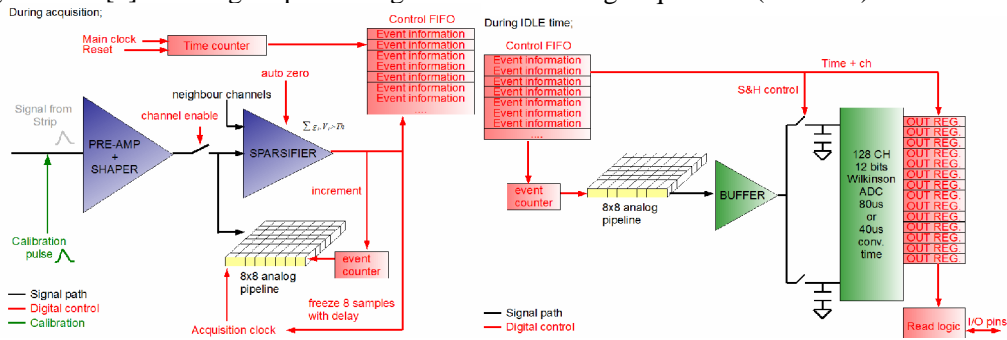

**Figure 4:** Schema of the data acquisition (left) and idle (right) periods as defined in the FE readout chip

All the channels in the chip are converted in parallel during the IDLE time. There are 8x8 samples to be converted per channel with a conversion time per channel of 85μs. The acquisition is based on a simple finite state machine with 4 states: IDLE (nothing to do, waiting for starting DAQ, time for powering off), START-PIPE, (pipelines initialization), WRITE (after initializing the pipelines, WRITE can start, triggered by the sparsifier if a detected signal), READ (a cycle of conversion/ read is repeated until no more data).

### 3.2 Level 2-DAQ

At the periphery of the detectors DSPs are foreseen for gathering the information from several daisy chained chips. They will be used for buffering, data compression and eventually further digital processing and interface with the outside world i.e. to send the data to the central DAQ.

### 3.3 Level 3-DAQ

The level 3 DAQ is located in the control room. It masters all the needed control signals to be properly distributed. It can perform more sophisticated processing (ex. Track reconstruction). It ensures the proper merging of Silicon tracking data with the other sub-detectors and the central DAQ. Whereas the level 1 and partly the level 2 are based on full custom devices, the rest of the DAQ will be mostly based on up-to-date products available in the market.

## 4 First steps for developing the Si tracking DAQ architecture

The SiLC test beam activities started in 2006 at DESY. Since 2007, a DAQ system is being developed for tests at CERN within the framework of the EUDET E.U. Project [1]. A first fully standalone Silicon tracking system was developed in 2008 [5] with DAQ system adapted to the new SiTR_130-88 chips and thus constitutes a first embryo of the future DAQ (Fig.5).

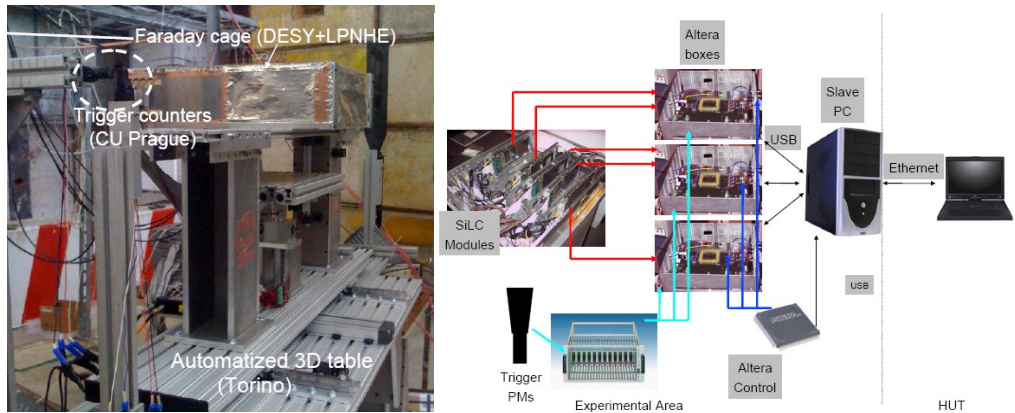

**Figure 5:** Standalone Si system test beam set-up at CERN (left) and synopsis of the related DAQ (right)

## 5    Concluding remarks

The FE readout chip on detector is a crucial piece of the Silicon DAQ with a high level of data processing and handling. It is the Level 1 DAQ central piece and the related R&D project is advancing well. The Level 2 DAQ is located on detector with a DSP-like interface to the outside world. The DAQ Level 3 is the master managing all the controls and combining Silicon data with the other sub-detector data. Real time reconstruction of tracks is considered at this Level. Close contacts with Industry are essential to avoid useless and expensive R&D work and a system soon obsolete. Cabling and data transmission/processing, evolving very fast, have to be seriously followed. Test beams are essential in the development of the DAQ.